# A Proof System for a Logic of Presuppositions

*X.Y. Newberry*

**Abstract**

The paper proposes a derivation system for a logic of presuppositions as introduced by P. F. Strawson. It is based on truth-relevant logic described by M. Richard Diaz in 1981. In another paper I outlined a derivation system for t-relevant logic based on truth trees. The conclusion was that a tautology is truth-relevant iff all the variables in a tree are self-contradicted. It is possible that a tree terminates without all the variables self-contradicting themselves. In this case the pertaining formula is still a tautology, but not a truth-relevant tautology. This concept is extended to the predicate calculus, i.e. a logic of presuppositions or a variant thereof.

# 1. Introduction

The purpose of this paper is to establish a derivation system for a logic where vacuous sentences, such as ∀*x(x ≠ x → x = x+1)*, are not derivable. In such a logic Gödel's self-referential sentence is not derivable [it is vacuous], but *~(Ex)Prf(x, g)* might be. Here *g* is the Gödel number of Gödel's sentence. And *Prf(x, z)* means that *x* is the Gödel number of a sequence that is a proof of the sentence with Gödel number *z*. (Newberry, 2015)

In Newberry (2019b) I suggested to use truth-relevant logic of M. Richard Diaz (1981) as a basis for this system. I proposed a derivation system for this logic logic based on tableaux. (The tableau method is described e.g. in Bergman at al, 1980.) The main conclusion was that a tautology *L* is t-relevant if all the propositional variables occurring in *L* are self-contradicted. This paper attempts to extend the same methods to the predicate calculus. We will accomplish this by "translating" predicate sentences into propositional sentences.

# 2. "Translation" to Predicate Calculus

## 2.1 Unquantified Sentences with Constants

The case when a sentence contains only predicates applied to to constants, i.e. sentences without variables and quantifiers, is straightforward. Predicates applied to a constant such as *Pa*, *Pb*, *Qa*, *Qb* behave the same way as distinct propositional variable of propositional logic. When all of them are self-contradicted the sentence is truth-relevant Example 2.1 in Appendix A.

## 2.2 Unquantified Sentences with Identity

Identity will group the constants into equivalence classes. That is given *a = b = c* means that *a*, *b*, *c* are naming the same object. Then *Pa*, *Pb*, *Pc* will count as one and the same atomic sentence. We could possibly have a sentence a = b v a = d. This implies that each branch will have its own equivalence classes.

As a practical matter as a first step in a proof we "consolidate" all constants in an equivalence class into one. For example we replace *b* and *c* with *a*. Then the same principles will apply as in section 2.1 above.

## 2.3 Universally Quantified Sentences

This case is also fairly straightforward. A sentence $\exists x Ux$ will get converted into a universally quantified sentence at the top of a tableau. That is it becomes **~$\exists x Ux$**, i.e. **$\forall x \sim Ux$**. But the latter will get instantiated as $\sim Ua$ for some $a$. If there are any constants in the sentence then $x$ will be instantiated as one of these constants. We thus obtain predicates with constants, and they behave as in section 2.1 above. Example 2.2.

## 2.4 Existentially Quantified Sentences

This case is a little bit more complicated. A sentence $\forall x Ux$ will get converted into an universally quantified sentence at the top of a tableau. That is it becomes **~$\forall x Ux$**, i.e. **$\exists x \sim Ux$**. It will get instantiated as $\sim Ue$ for some name $e$. Now the problem is that although e is different from all the constants in the tree we do not know if $e$ names any object already mentioned in the tree. How do we analyze this situation?

Consider this example

$(P_1c_1 \ \& \sim P_1c_1) \lor (P_2c_2 \ \& \sim P_2c_2) \lor \ldots (P_ic_i \ \& \sim P_ic_i) \lor (\mathbf{P_2e} \ \& \sim \mathbf{P_2e}) \lor \ldots$   (2.4.1)

What if $e$ names the same object as $c_2$ for example? Does it mean that **$P_2e$** is redundant even though (2.4.1) indicates that it is not? The answer is that $e$ does not stand for any particular object; $P_2e$ is more like a schema. The subformula $P_2e$ is the result of instantiating some existential sentence, say $\exists x P_2 x$. The sentence $\exists x P_2 x$ assures us that there is at least one object $e$ such that $P_2e$, but there is no indication which object it might be. So the question we need to ask is: how can we prove anything at all if we do not know which object we are talking about. If you look at the structure of the proof it actually works with *any object e* in the domain. This is because (if the tree closes) the existential quantifier is invariably met with a universal quantifier, which casts it nets really wide. Instantiate $\exists x P_2 x$ with any name, and the waiting universal quantifier will instantiate with the very same name! (Example 2.3, line 9.) Do not try this at home, but you could even instantiate with $c_1$, and the tree would still close. Again, $e$ does not stand for any particular object; it can be interpreted as any object. The proof has to work even if $e$ is interpreted as an object not identical with $c_2$. We are therefore justified regarding $P_2e$ as relevant.

But now suppose we obtain

$(P_1c_1 \ \& \sim P_1c_1) \lor (P_2c_2 \ \& \sim P_2c_2) \lor \ldots (P_ic_i \ \& \sim P_ic_i) \ \& \ \mathbf{P_ie} \lor \ldots$   (2.4.1)

or

$(P_1c_1 \ \& \sim P_1c_1) \lor (P_2c_2 \ \& \sim P_2c_2) \lor \ldots (P_ic_i \ \& \sim P_ic_i) \ \& \ (\mathbf{P_ie} \ \& \sim \mathbf{P_ie}) \lor \ldots$   (2.4.2)

This would happen for example if ∃xPx and Pc, for some constant c, appeared on the same branch.

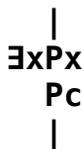

Figure 2.4.1

The formulas (2.4.1) and (2.4.2) indicate that $P_i e$ is redundant. Here ∃xPx is a consequence of Pc, and it has a wider meaning than Pc; it is indeed redundant.

## 2.5 Quantified Sentences with Equivalence

We observe that

1) Equivalence groups constants into equivalence classes; each branch has its own equivalence classes.

2) The formulas $c = x$, $c = y$ etc. for any constant c are just other predicates;  if $a \neq b$ then $a = x$ and $b = x$ are two different predicates.

3) We can view $x \neq x$ as $x = c$ & $x \neq c$ for some constant c.

Example 2.4.

## 2.6 Polyadic Sentences

We better first look at some examples.

**~(∃x)(∃y)(Fxy & Gy)**

Let *Fxy =def=* (x + y <= 8), *Gy =def=* (y = 12). We obtain ~∃x∃y(x + y <= 8 & y = 12).
This is Gödel-like sentence. According to Figure 7.1 below it is not t-relevant.

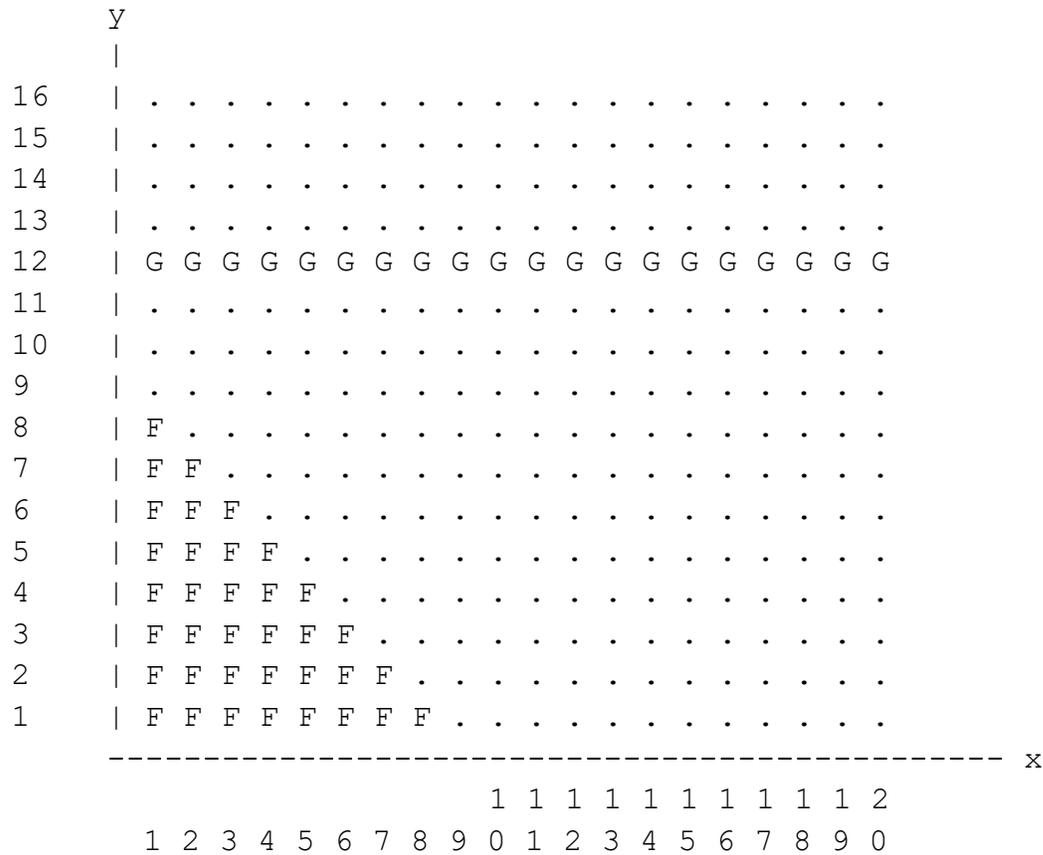

Figure 7.1

Let us apply the tableau method:

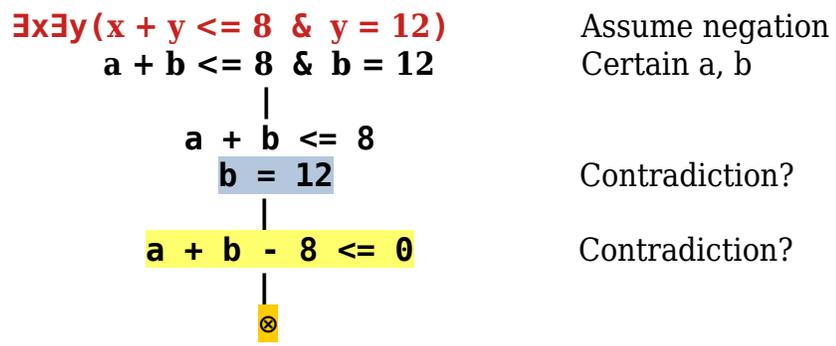

There are two potential contradictions and *only one branch*! (The branch will close as soon as the first contradiction is encountered.) Both predicates cannot be relevant.

Suppose we fix *y* at **12**: ~∃x∃y(x + **12** <= 8 & **12** = 12)

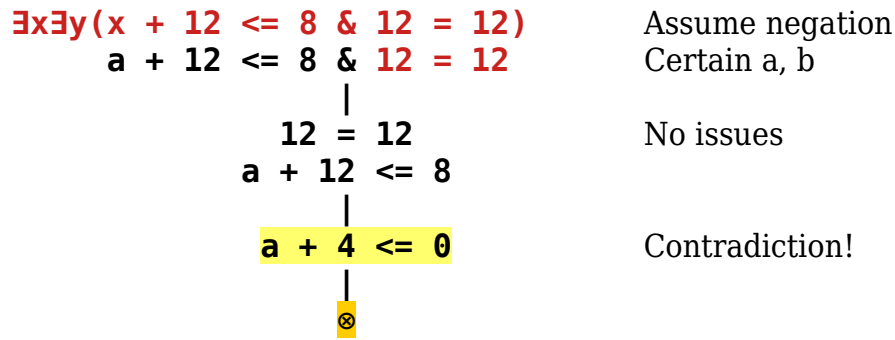

The predicate *12 = 12* is redundant.

Suppose we fix *y* at **7**: ~∃x∃y(x + **7** <= 8 & **7** = 12)

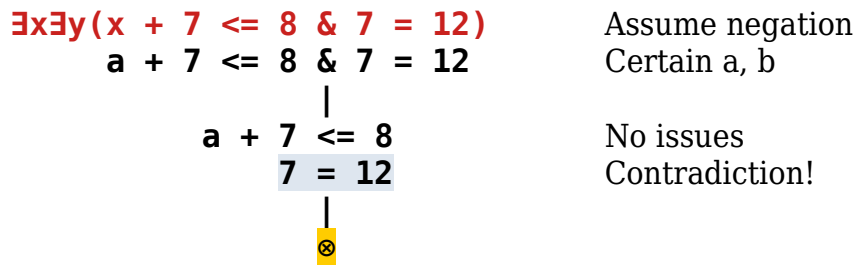

The predicate *x + 7 <= 8* is redundant.

Suppose we fix *y* at **9**: *~(∃x)(∃y)(x + **9** <= 8 & **9** = 12)*

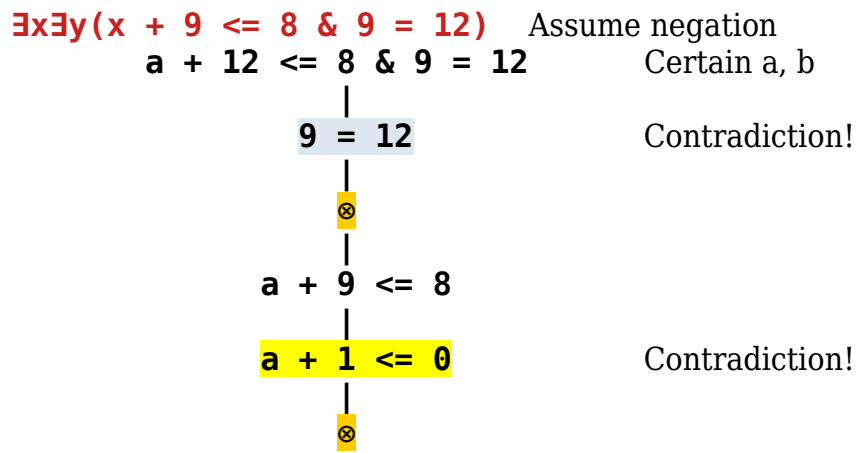

The branch will close when whichever contradiction occurs first. The formula has two t-relevant subsets, namely *{x + 9 <= 8}* and *{9 = 12}*.

But for no natural number n will the formula *~(∃x)(∃y)(x + **n** <= 8 & **n** = 12)* be t-relevant. Therefore *~∃y(∃x(x + y <= 8) & ∃x(y = 12)* is *not* t-relevant.

The sentence *~(∃x)(∃z)(x + z <= 8 & z = 12)* bears a strong similarity to Gödel's sentence *G = ~(∃x)(∃z)(Prf(x,z) & Diag(k,z))*, where *Prf()* means that that *x* is (the Gödel number of) a sequence that is a proof of the sentence (with Gödel number) *z*, and *Diag(k,z)) is satisfied only by* the Gödel number ⌜G⌝ of *G*. If it is indeed the case that *~(∃x)(Prf(x,⌜G⌝)* then graphically the situation will look much like Figure 7.1 in the sense that the sets of pairs *(x,z)* of natural numbers picked by *Prf()* and *Diag()* respectively will not intersect. That means that *G* will *not* be t-relevant even if *~(∃x)(Prf(x,⌜G⌝)* is true and provable!

* * * * * * *

Compare the above to *Fxy =def= (x + y < 8), Gy =def= (x + y > 8)*. We obtain ~∃x∃y(x + y < 8 & x + y > 8).

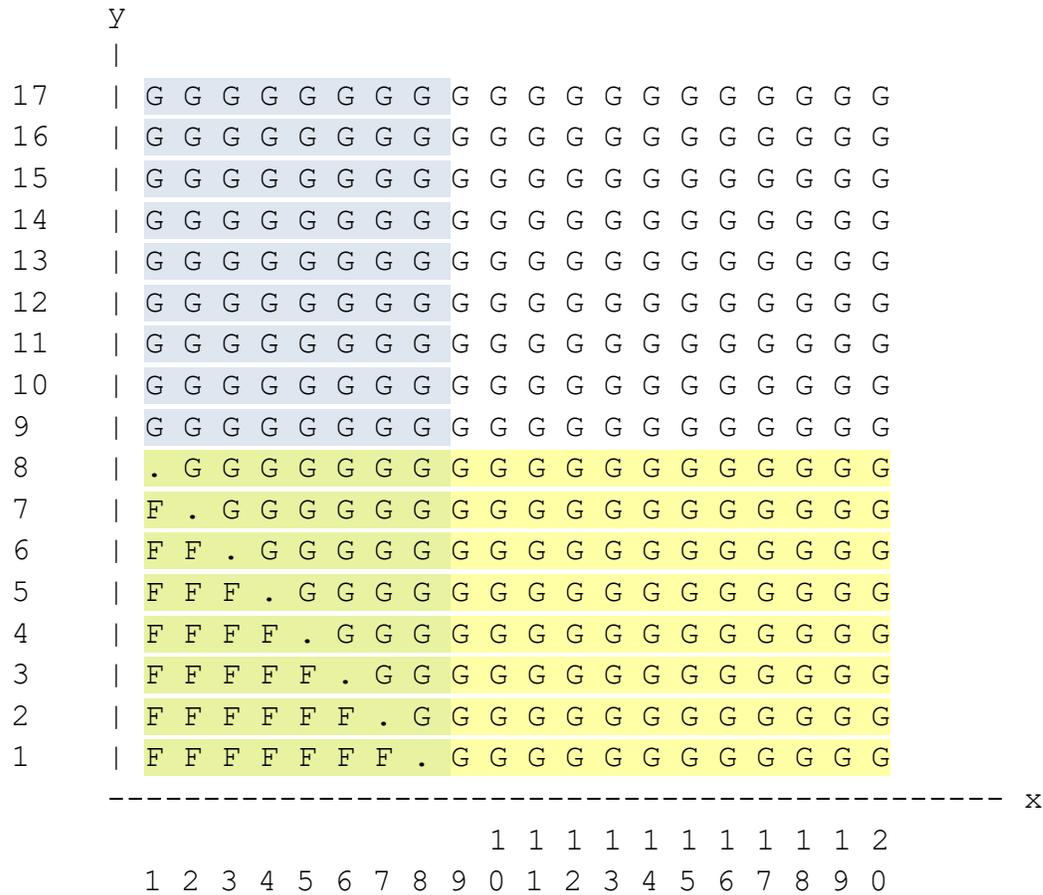

Figure 6.2

Here is a tree.

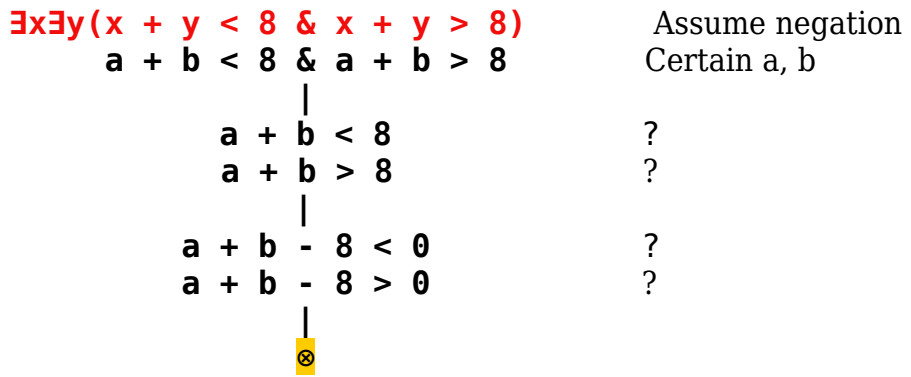

There are numbers *a* and *b* such that *a + b < 8*, and there are numbers *a* and *b* such that *a + b > 8*. No issues. The predicates are not self-contradictory in themselves.

Nevertheless they do contradict each other. Both are required for a contradiction to occur, both are relevant!

When we fix *y* anywhere in the interval [1,7] we obtain a t-relevant sentence. For example let y = 3:

```
           ∃x(x + 3 < 8  &  x + 3 > 8)        Assume negation
             a + 3 < 8  &  a + 3 > 8          Certain a, b
                         |
                  a + 3 < 8                   No contradiction
                  a + 3 > 8                   No contradiction
                         |
                  a + 3 - 8 < 0               No contradiction
                  a + 3 - 8 > 0               No contradiction
                         |
                         ⊗
```

Nevertheless  *a + 3 < 8* and  *a + 3 > 8* contradict each other as there is no natural number that satisfies both. When we fix *y* anywhere in the interval [8,∞) we obtain a non-relevant sentence:

```
           ∃x(x + 9 < 8  &  x + 9 > 8)        Assume negation
             a + 3 < 8  &  a + 3 > 8          Certain a, b
                         |
                  a + 9 < 8                   Contradiction brewing
                  a + 9 > 8                   No contradiction
                         |
                  a + 1 < 0                   Contradiction!
                  a + 1 > 0                   No contradiction
                         |
                         ⊗
```

The predicate *x + 9 > 8* is redundant. The situation is completely symmetric with respect to *x*.

# Appendix A – Examples

**Example 2.1 - Unquantified sentence with constants.**

*(Pa → Pb) → (~Pb x → ~Pa)*. It is a tautology and below is a tableau proof.

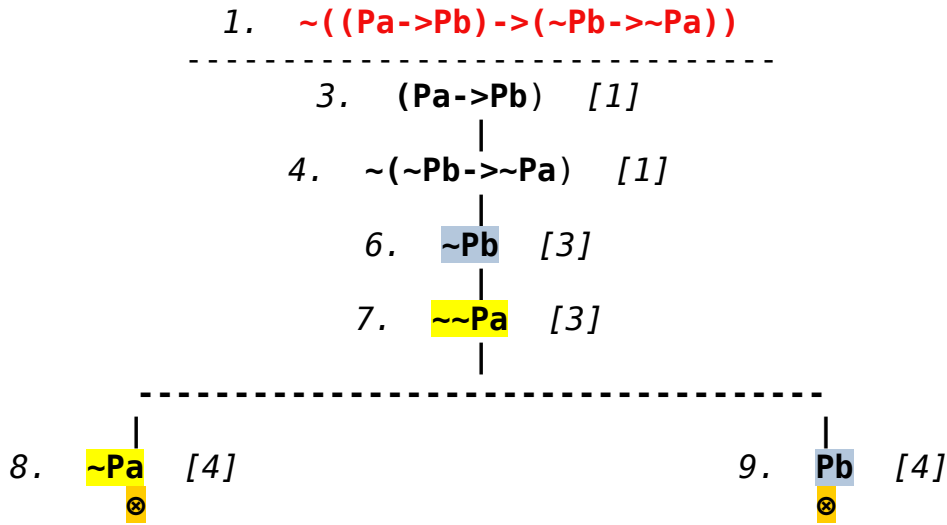

The predicates *Pa* and *Pb* behave like a propositional variables in propositional logic. They are both self-contradicted, i.e. the sentence is truth relevant and hence true.

**Example 2.2 - Universally quantified sentence**

Here is an example: *(Fm ∨ Fn), ∀x(Fx → Gx) ⊢ ∃xGx*

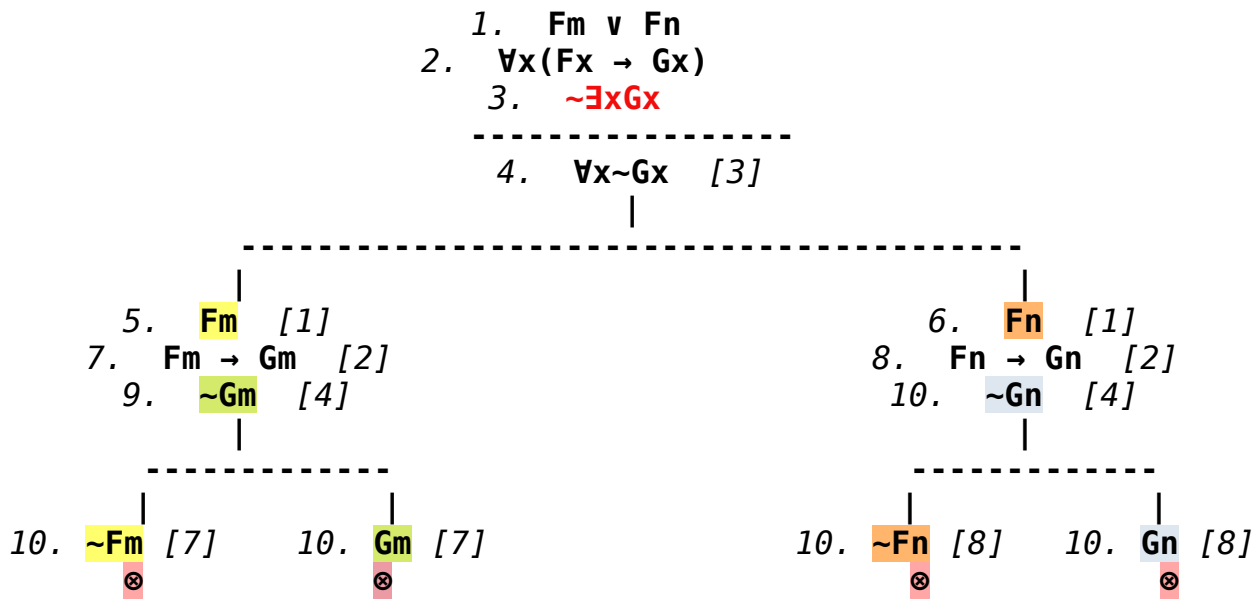

The resulting tree has four atomic predicates: *Fm, Fn, Gm, Gn*. All of these are self-contradicted, and we conclude that the implication *(Fm ∨ Fn) & ∀x(Fx → Gx) -> ∃xGx* is valid and t-relevant, i.e. true.

We can in fact "translate" the set of the three above sentences into a set of unquantified sentences by exhibiting a counterexample:

*(Fm ∨ Fn), (Fm → Gm) & (Fn → Gn), ~(Gm ∨ Gn).*

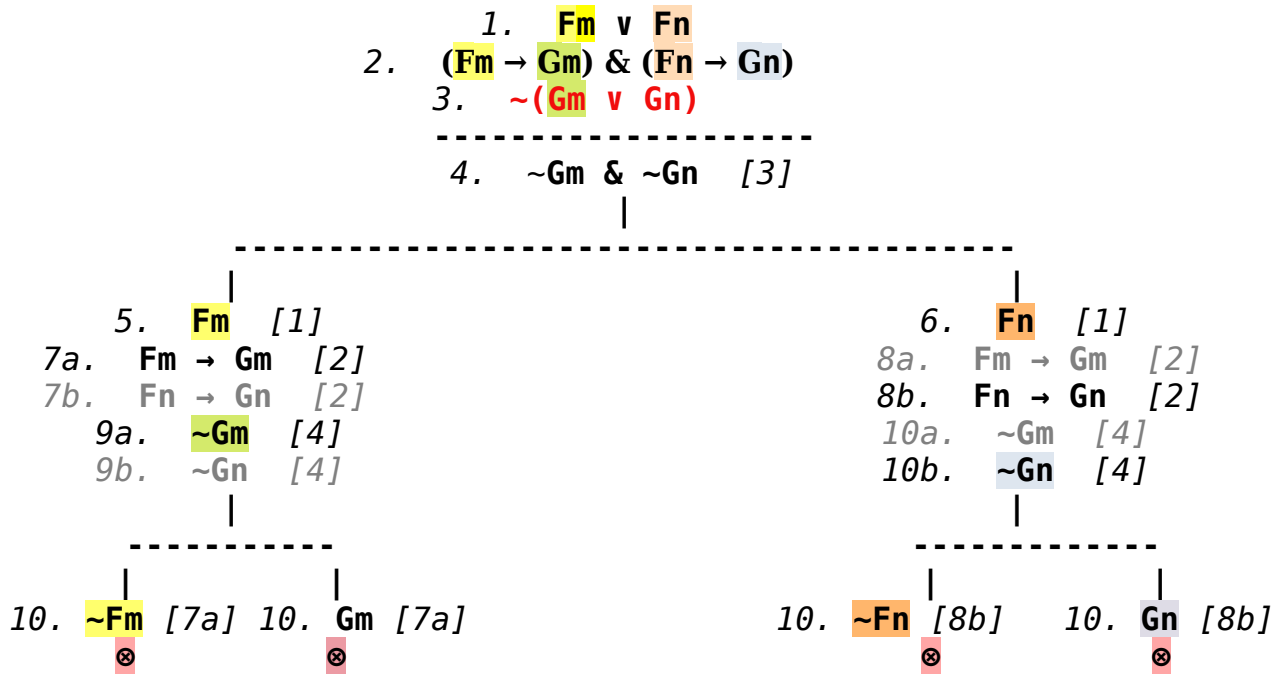

## Example 2.3 - Existentially quantified sentence

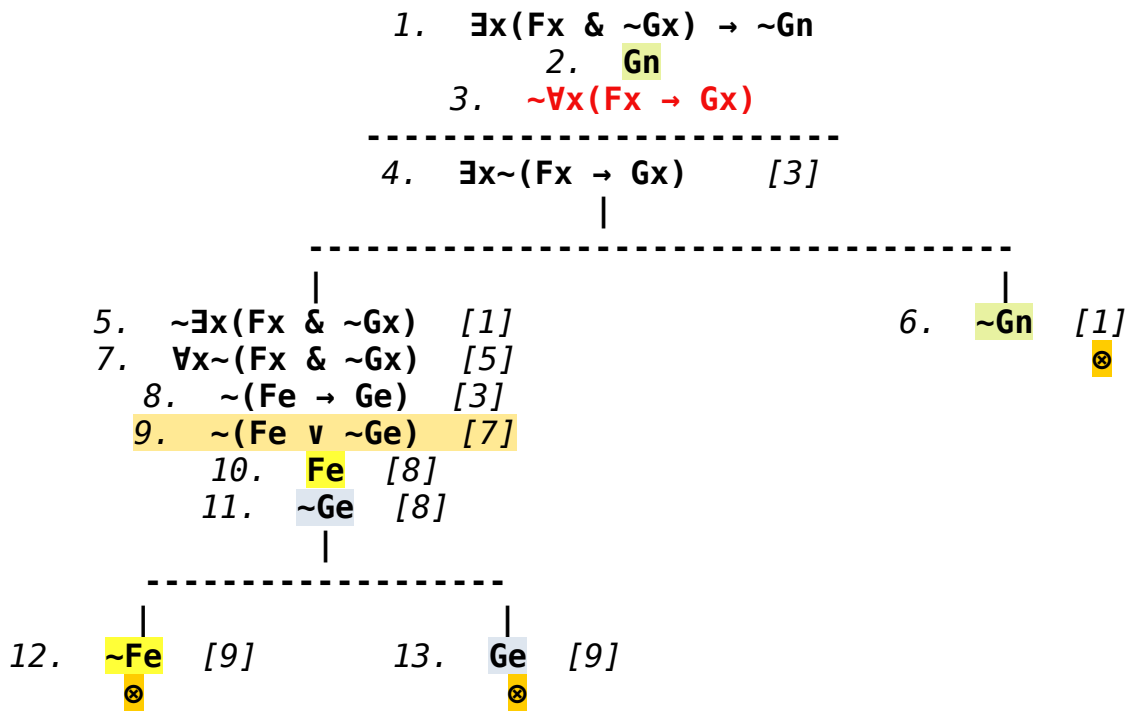

We can "translate" the above premises into unquantified sentences.

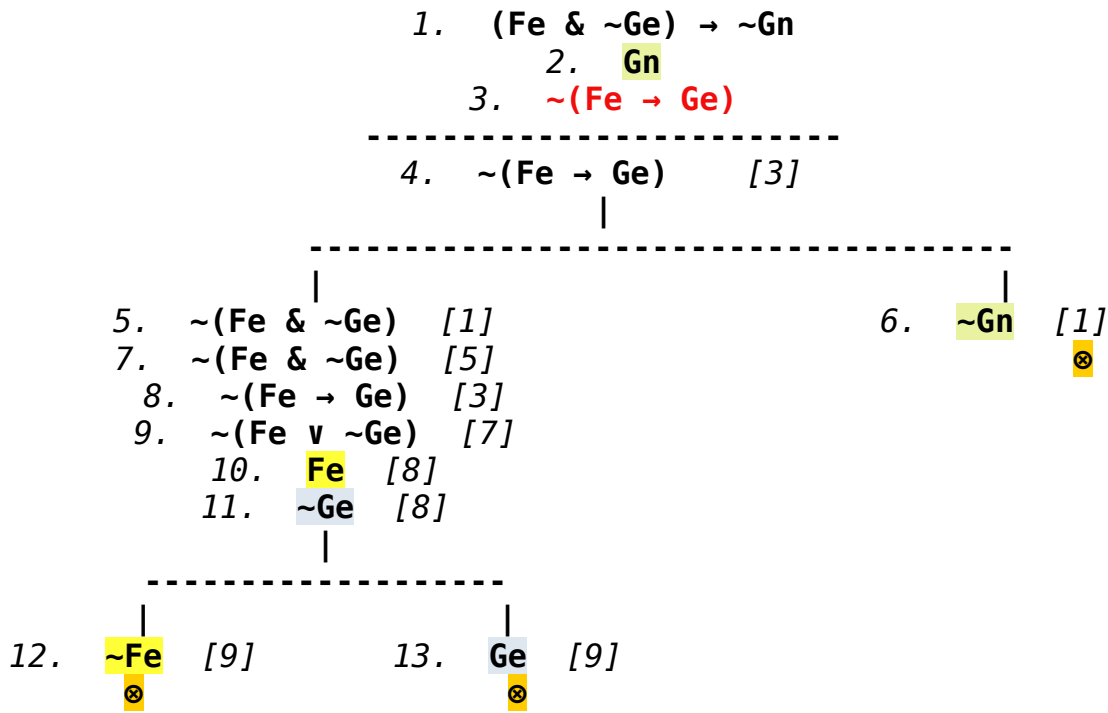

Here *e* can be interpreted as any object, even *n*, and the tree will still close. But the proof has to work even if *e* is *not* interpreted as *n*. Therefore *Ge is* relevant.

**Example 2.4 – Identity**

*Fm & ∀x(Fx → x = m), Fn & Gn ⊢ Gm.*

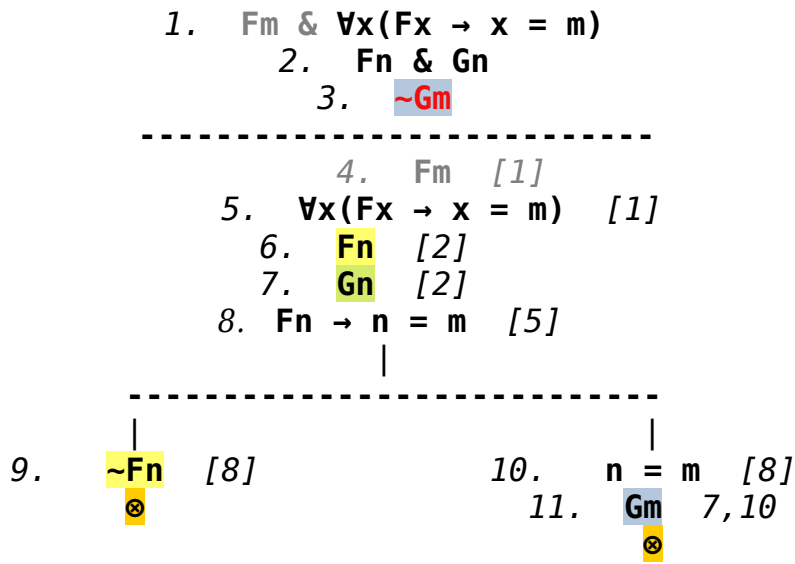

We have four primitive predicates: *Fm, Fn, Gm, Gn.* On the right branch we have obtained *n = m* as well as *Gm* self-contradiction; we do not need explicit *Gn* self-contradiction. Both *m* and *n* belong to the same equivalence class. On the left branch we have *Fn* self-contradiction but *no **n = m***, and no *Fm* self-contradiction. We can readily verify that the tree will work just as well if we simply leave *Fm* out. The same cannot be said about *Gn* for we have obtained *Gm* on line 11 from *Gn* on line 7. So *Fm* is *not* relevant.

# Appendix B – Prototypical Cases

Let *F* and *G* be atomic predicates. There are four prototypical cases.

**Case 1:** ~∃xKx  [∀x~Kx]

   Case 1.1: ~∃x(Fx & Gx) [∀x(~Fx v ~Gx)]

   Case 1.2: ~∃x(Fx v Gx) [∀x(~Fx & ~Gx)]   *clean board*

**Case 2:** ∃xKx

   Case 2.1: ∃x(Fx & Gx)                         *negation of 1.1*

   Case 2.2: ∃x(Fx v Gx)                         *negation of 1.2, non-empty board*

**Case 1.1: ~∃x(Fx & Gx)** [∀x(~Fx v ~Gx)]

The *presuppositions* are: ∃xFx and ∃xGx (Newberry 2019a.)

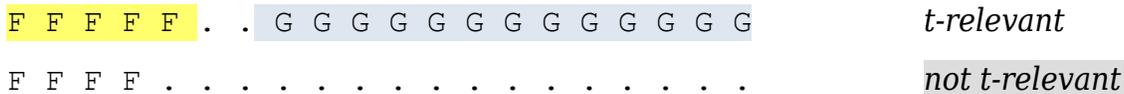   *t-relevant*

                                                *not t-relevant*

*F* and *G* do not overlap. Every object is *(~Fx v ~Gx)*.

By flipping the signs of *Fx*, *Gx*, i.e. *Jx =def= ~Fx, Kx =def= ~Gx* we obtain

*~∃x(~Jx & ~Kx) = ∀x(Jx v Kx)*, which is to say the last formula is not really a separate case.

Conditions of t-relevance (presuppositions): ∃x(~Jx) • ∃x(~Kx)

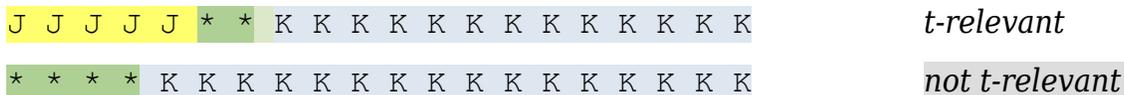

(The symbol '*' means both *J* and *K*.)

Example B1.1: ~∃x(x < 5 & x > 5)

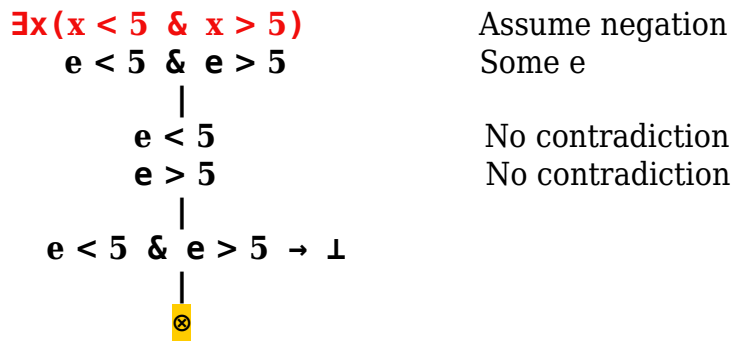

The two predicates contradict each other.

Example B1.2: ~∃x(x < 5 & x = x+1)

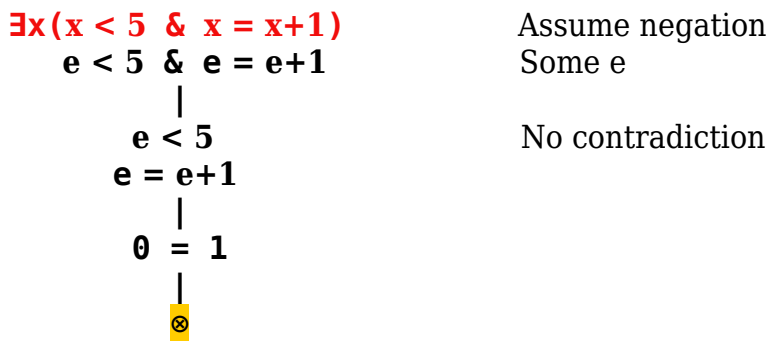

```
        ∃x(x < 5 & x = x+1)         Assume negation
          e < 5 & e = e+1           Some e
                 |
              e < 5                 No contradiction
              e = e+1
                 |
               0 = 1
                 |
                 ⊗
```

The predicate *x < 5* is redundant.

Example B1.3: ∀x(x >= 5 v x <= 5)

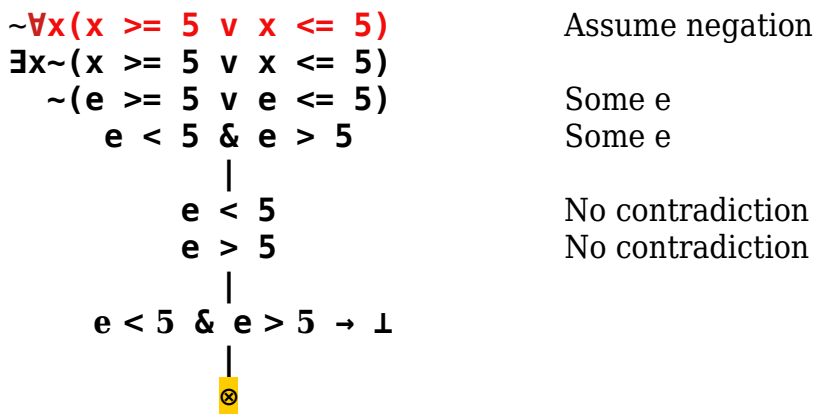

```
        ~∀x(x >= 5 v x <= 5)        Assume negation
         ∃x~(x >= 5 v x <= 5)
           ~(e >= 5 v e <= 5)       Some e
              e < 5 & e > 5         Some e
                   |
                 e < 5              No contradiction
                 e > 5              No contradiction
                   |
            e < 5 & e > 5 → ⊥
                   |
                   ⊗
```

The tree is very similar to case 1.1.1. Both predicates are relevant.

**Case 1.2: ~∃x(Fx v Gx)** [∀x(~Fx & ~Gx)]                    *clean board*

Both predicates must be empty for the sentence to hold.

. . . . . . . . . . . . . . . . . .

Case 1.2.1  ~∃x(x = x+1 v x = x+2)

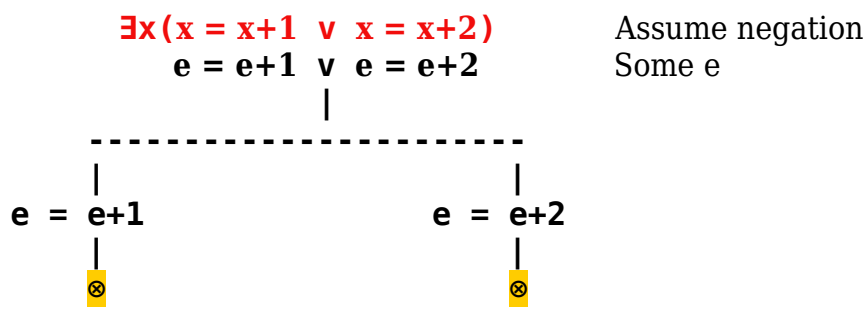

**Case 2.1: ∃x(Fx & Gx)**                                    *negation of 1.1*

Presuppositions are the same as in case 1.1: ∃xFx and ∃xGx. They are satisfied automatically if the sentence is satisfied.

F F F F F * * F F F G G G G G G G G

Example 2.1.1: ∃x(x <= 5 & x >= 5):

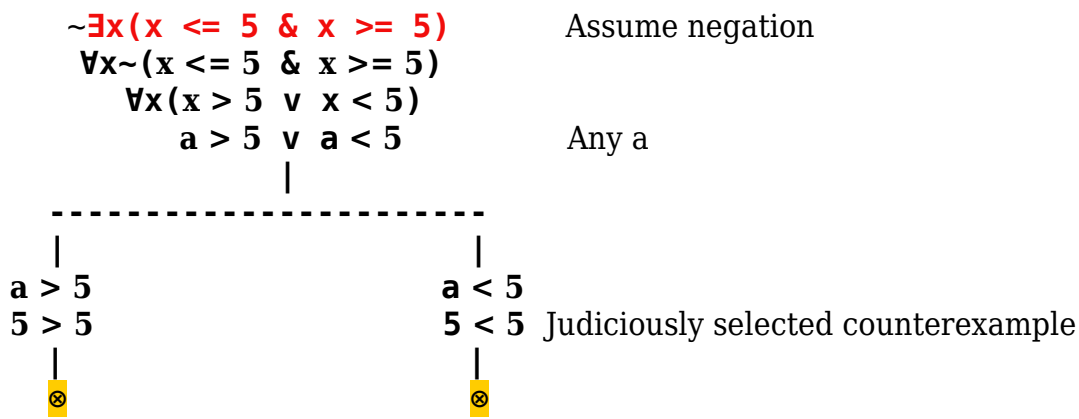

**Case 2.2: ∃x(Fx v Gx)**                              *negation of 1.2, non-empty board*

```
F F F F F . . G G G G G G G G G G G G
F F F F . . . . . . . . . . . . . . .
```

Case 2.2.1: ∃x(x < 5 v x > 5)

$$
\begin{array}{ll}
\sim\exists x(x < 5 \;v\; x > 5) & \text{Assume negation} \\
\forall x\sim(x < 5 \;v\; x > 5) & \\
\forall x(x >= 5 \;\&\; x <= 5) & \\
\quad a >= 5 \;\&\; a <= 5 & \text{Any a} \\
\quad\quad | & \\
\quad\quad a >= 5 & \\
\quad\quad 4 >= 5 & \text{Judiciously selected counterexample} \\
\quad\quad \otimes & \\
\quad\quad | & \\
\quad\quad a <= 5 & \\
\quad\quad 6 <= 5 & \text{Judiciously selected counterexample} \\
\quad\quad \otimes &
\end{array}
$$

Not t-relevant. The sentence has two t-relevant proper subsets: {x < 5} and {x > 5}.

Case 2.2.2: ∃x(x < 5 v x = x+1)

$$
\begin{array}{ll}
\sim\exists x(x < 5 \;v\; x = x+1) & \text{Assume negation} \\
\forall x\sim(x < 5 \;v\; x = x+1) & \\
\forall x(x >= 5 \;\&\; x \neq x+1) & \\
\quad a >= 5 \;\&\; a \neq a+1 & \text{Any a} \\
\quad\quad | & \\
\quad\quad a >= 5 & \\
\quad\quad a \neq a+1 & \text{Judiciously selected counterexample} \\
\quad\quad | & \\
\quad\quad 4 >= 5 & \text{Judiciously selected counterexample} \\
\quad\quad \otimes &
\end{array}
$$

The predicate x = x+1 is irrelevant.